\documentclass[preprint,preprintnumbers,amsmath,amssymb]{revtex4}
\usepackage{graphicx}
\usepackage{color}

\def\beq{\begin{equation}}
\def\eeq{\end{equation}}
\def\eeqn{\end{equation}}
\newcommand\iden{\leavevmode\hbox{\small1\normalsize\kern-.33em1}}

\newcommand{\bea} {\begin{eqnarray}}
\newcommand{\eea} {\end{eqnarray}}

\def\tb {t_\beta}
\def\sb  {s_{\beta}}
\def\cb  {c_{\beta}}

\def\lam{\lambda}

\let\jnfont=\rm
\def\NPB#1,{{\jnfont Nucl.\ Phys.\ B }{\bf #1},}
\def\PLB#1,{{\jnfont Phys.\ Lett.\ B }{\bf #1},}
\def\EPJC#1,{{\jnfont Eur.\ Phys.\ Jour.\ C }{\bf #1},}
\def\PRD#1,{{\jnfont Phys.\ Rev.\ D }{\bf #1},}
\def\PRL#1,{{\jnfont Phys.\ Rev.\ Lett.\ }{\bf #1},}
\def\MPLA#1,{{\jnfont Mod.\ Phys.\ Lett.\ A }{\bf #1},}
\def\JPG#1,{{\jnfont J.\ Phys.\ G }{\bf #1},}
\def\CTP#1,{{\jnfont Commun.\ Theor.\ Phys.\ }{\bf #1},}
\def\JHEP#1,{{\jnfont JHEP \ }{\bf #1},}
\def\NPPS#1,{{\jnfont Nucl.\ Phys.\ Proc.\ Suppl.\ }{\bf #1},}
\def\CPC#1,{{\jnfont Computl.\ Phys.\ Commun.\ }{\bf #1},}
\def\CPL#1,{{\jnfont Chin.\ Phys.\ Lett. }{\bf #1},}
\def\AJS#1,{{\jnfont Astrophys.\ J.\ Suppl. }{\bf #1},}
\def\PR#1,{{\jnfont Phys.\ Rept. }{\bf #1},}
\def\AP#1,{{\jnfont Astropart.\ Phys. }{\bf #1},}
\def\EPL#1,{{\jnfont Europhys.\ Lett. }{\bf #1},}
\def\FP#1,{{\jnfont Fortsch.\ Phys. }{\bf #1},}
\def\JCAP#1 {{\jnfont JCAP \ }{\bf #1} }

\begin{document}

\title{Electroweak baryogenesis, dark matter, and dark CP-symmetry}
\renewcommand{\thefootnote}{\fnsymbol{footnote}}

\author{Jinghong Ma, Jie Wang, Lei Wang$^{*}$\footnotetext{*) 
Corresponding author. Email address: leiwang@ytu.edu.cn}}
 \affiliation{Department of Physics, Yantai University, Yantai
264005, P. R. China}
\renewcommand{\thefootnote}{\arabic{footnote}}

\begin{abstract}
With the motivation of explaining the dark matter and achieving the electroweak baryogenesis via the spontaneous CP-violation at high temperature,
we propose a complex singlet scalar ($S=\frac{\chi+i\eta_s}{\sqrt{2}}$) extension of the two-Higgs-doublet model respecting a discrete dark CP-symmetry: $S\to -S^*$.
The dark CP-symmetry guarantees $\chi$ to be a dark matter candidate on one hand and on the other hand allows $\eta_s$ to have mixings with 
the pseudoscalars of the Higgs doublet fields, which play key roles in generating the CP-violation sources needed by
the electroweak baryogenesis at high temperature. The universe undergoes multi-step phase transitions,
 including a strongly first-order electroweak phase transition during which the baryon number is produced.
At the present temperature, the observed vacuum is produced and the CP-symmetry is restored so that the
stringent electric dipole moment experimental bounds are satisfied. Considering relevant constraints, we study the simple scenario of
$m_{\chi}$ around the Higgs resonance region, and find that the dark matter relic abundance and the baryon asymmetry can be simultaneously explained.
Finally, we briefly discuss the gravitational wave signatures at future space-based detectors and the LHC signatures.
\end{abstract}
\maketitle

\section{Introduction} 
The baryon asymmetry of the universe (BAU) presents one of the major quests for particle
cosmology. By the observation based on the Big-Bang Nucleosynthesis, the BAU is \cite{pdg2020}
\beq
Y_B \equiv \rho_B/s = (8.2 - 9.2) \times 10^{-11},
\eeq  
 where $\rho_B$ is the baryon number density and $s$ is the entropy density.
Three necessary Sakharov conditions have to be fulfilled for a dynamical generation of BAU: baryon number changing interactions, 
non-conservation of C and CP, and departure from thermal equilibrium \cite{Sakharov}.
The electroweak baryogenesis (EWBG) \cite{ewbg1,ewbg2} provides a promising and
attractive mechanism of explaining the BAU since it is testable at the energy frontier by the LHC and at the precision frontier by the electric
dipole moment (EDM) experiments.
To realize the EWBG, one needs extend the SM to produce sufficient large CP-violation and a strongly first-order electroweak phase transition (EWPT),
such as the singlet extension of SM (see e.g. \cite{bgs-1,bgs-3,bgs-4,bgs-5,bgs-6,Beniwal:2018hyi,Huang:2018aja,Ghorbani:2017jls,cao,huang,Xie:2020wzn,Ellis:2022lft,Lewicki:2021pgr,Idegawa:2023bkh,Harigaya:2022ptp}) 
and the two-Higgs-doublet model (2HDM) (see e.g. \cite{bg2h-1,bg2h-2,bg2h-3,Kanemura:2004ch,Basler:2017uxn,Abe:2013qla,bg2h-5,bg2h-4,bg2h-6,bg2h-7,bg2h-8,bg2h-9,bg2h-11,Basler:2020nrq,bg2h-13,bg2h-12,2111.13079,2207.00060,Goncalves:2023svb}).

The negative results in the EDM searches for electrons impose stringent constraints on the explicit CP-violation interactions in the scalar
couplings and Yukawa couplings  \cite{edm-e}. 
There are some cancellation mechanisms of CP-violation effects in the EDM, which can relax the tension between the EWBG and the EDM data \cite{1411.6695,Fuyuto:2017ewj,Modak:2018csw,Fuyuto:2019svr,Modak:2020uyq,2004.03943,2111.13079,2207.00060}. Even with the cancellation, there are several CP
observables of radiative $B$ meson decays that still provide stringent constraints, such as the asymmetry of the CP-asymmetry
of inclusive $B\to X_s\gamma$ decay \cite{Modak:2018csw,Modak:2020uyq},
Alternatively, a finite temperature spontaneous CP-violation mechanism is naturally compatible with the EDM data, where the CP
symmetry is spontaneously broken at the high temperature and it is recovered at the present temperature. 
The novel mechanism was achieved in the singlet scalar extension of the SM \cite{cao,huang} in which a high dimension effective
operator needs to be added,
and the singlet pseudoscalar extension of 2HDM \cite{Huber:2022ndk,Liu:2023sey}.

In addition to the BAU, the dark matter (DM) is one of the longstanding questions of particle
physics and cosmology. In this paper, we propose a complex singlet scalar extension of the 2HDM respecting a discrete dark CP-symmetry, 
and simultaneously explain the observed DM relic density and the BAU via the spontaneous CP-violation at high temperature. The dark CP-symmetry allows the imaginary component of singlet scalar
 to have mixings with the pseudoscalars of scalar doublet fields, which play key roles in generating the CP-violation sources needed by
the EWBG at high temperature. On the other hand, the dark CP-symmetry guarantees
 the real component to be a DM candidate.


\section{2HDM + $S$ respecting a dark CP-symmetry} 
Imposing a discrete dark CP-symmetry, we extend the SM by a second Higgs doublet $\Phi_2$ and a complex singlet $S$,
\bea
&&\Phi_1=\left(\begin{array}{c} \phi_1^+ \\
\frac{(v_1+\rho_1+i\eta_1)}{\sqrt{2}}\,
\end{array}\right)\,, 
\Phi_2=\left(\begin{array}{c} \phi_2^+ \\
\frac{(v_2+\rho_2+i\eta_2)}{\sqrt{2}}\,
\end{array}\right),
S=\frac{(\chi+i\eta_s)}{\sqrt{2}},
\eea
with $v_1$ and $v_2$ being the vacuum expectation values (VEVs), $v = \sqrt{v^2_1 + v^2_2} = (246~\rm GeV)^2$ and  $\tan\beta \equiv v_2 /v_1$. 
The singlet field $S$ has no VEV.
Under the dark CP-symmetry, $S\to-~S^*$ ($\chi \to -\chi$ ,$\eta_s \to \eta_s$ in the real parametrization), and
while all the other fields are not affected.

The full scalar potential is given as
\begin{eqnarray} \label{V2HDM} &&\mathrm{V} = m_{11}^2
(\Phi_1^{\dagger} \Phi_1) + m_{22}^2 (\Phi_2^{\dagger}\Phi_2) + \frac{\lambda_1}{2}  (\Phi_1^{\dagger} \Phi_1)^2 +
\frac{\lambda_2}{2} (\Phi_2^{\dagger} \Phi_2)^2 + \lambda_3
(\Phi_1^{\dagger} \Phi_1)(\Phi_2^{\dagger} \Phi_2) \nonumber \\
&& + \lambda_4
(\Phi_1^{\dagger}
\Phi_2)(\Phi_2^{\dagger} \Phi_1)+ \left[\frac{\lambda_5}{2} (\Phi_1^{\dagger} \Phi_2)^2  +  \lambda_6 (\Phi_1^{\dagger} \Phi_1) (\Phi_1^{\dagger} \Phi_2) + \lambda_7 (\Phi_2^{\dagger} \Phi_2) (\Phi_1^{\dagger} \Phi_2)  +\rm
h.c.\right]\nonumber\\
&&+m^2_{S}SS^* + \left[\frac{m^{\prime 2}_{S}}{2}SS+\rm h.c. \right] + \left[\frac{\lambda^{\prime\prime}_1}{24}S^4+\rm h.c. \right] 
+\left[\frac{\lambda^{\prime\prime}_2}{6}S^2SS^*+\rm h.c. \right] \nonumber\\
&&+SS^*\left[\lambda^{\prime}_1\Phi_1^{\dagger} \Phi_1+\lambda^{\prime}_2\Phi_2^{\dagger} \Phi_2 \right]
+\frac{\lambda^{\prime\prime}_3}{4}(SS^*)^2+\left[S^2 (\lambda^{\prime}_4\Phi_1^{\dagger} \Phi_1+\lambda^{\prime}_5\Phi_2^{\dagger} \Phi_2) +\rm h.c. \right]\nonumber\\
&&+ \left[\lambda^{\prime}_6 SS^* \Phi_2^{\dagger} \Phi_1 + \lambda^{\prime}_7 (SS+S^*S^*) \Phi_2^{\dagger} \Phi_1 + \rm h.c. \right]\nonumber\\
&&+ \left[-m_{12}^2 \Phi_2^{\dagger} \Phi_1 + \frac{\mu}{2} (S-S^*) \Phi_1^{\dagger} \Phi_2  + \rm h.c. \right],
\end{eqnarray}
where all the coupling coefficients and mass are real, and thus the scalar potential sector is CP-conserved at zero temperature.
The last term leads to the mixings of $\eta_s$ and the pseudoscalars of Higgs doublet fields, and $\chi$ is allowed to remain stable.
For simplicity, we take $\lambda_6=\lambda_7=\lambda^\prime_6=\lambda^\prime_7=\lambda^{\prime\prime}_2=0$ in the following discussions.

The stationary conditions give
\bea
&&\quad m_{11}^2 = m_{12}^2 \tb - \frac{1}{2} v^2 \left( \lambda_1 \cb^2 + \lambda_{345}\sb^2 \right)\,,\nonumber\\
&& \quad m_{22}^2 =  m_{12}^2 / \tb - \frac{1}{2} v^2 \left( \lambda_2 \sb^2 + \lambda_{345}\cb^2 \right),
\label{min_cond}
\eea
where $\tb\equiv \tan\beta$, $\sb\equiv \sin\beta$, $\cb \equiv \cos\beta$,
and $\lambda_{345} = \lambda_3+\lambda_4+\lambda_5$.

In addition to the 125 GeV Higgs $h$, the physical scalar spectrum contains a CP-even states $H$, a DM candidate $\chi$,
 two neutral pseudoscalars $A$ and $X$, and a charged scalar $H^{\pm}$. 
The mass eigenstates $h$, $H$ and $H^{\pm}$ and their masses are the same as those of the pure 2HDM. 
The $\eta_1$, $\eta_2$ and $\eta_s$ are rotated into the $A$, $X$ and $G$ by the two mixing angles $\theta$ and $\beta$,
where $G$ is a Goldstone boson.
The parameters $\mu$, $m_S^2$, and $m_S^{\prime 2}$ are given as
\bea
&& m_S^2 =\frac{1}{2} \left(m_\chi^2+m_A^2 s_\theta^2+ m_X^2 c_\theta^2  -\lambda^\prime_1 v^2 \cb^2 - \lambda^\prime_2 v^2 \sb^2\right),\nonumber\\
&& m^{\prime 2}_S =\frac{1}{2} \left(m_\chi^2- m_A^2 s_\theta^2 - m_X^2 c_\theta^2  -2\lambda^\prime_4 v^2 \cb^2 - 2\lambda^\prime_5 v^2 \sb^2\right),\nonumber\\
&& \mu = \frac{\sqrt{2}(m_{X}^2-m_A^2)}{v}s_\theta c_\theta,
\label{eq:mum0}
\eea
where $s_\theta\equiv \sin\theta$ and $c_\theta \equiv \cos\theta$.
The couplings $\lambda_i$ ($i=1,2,3,4,5$) are determined by
\begin{eqnarray}\label{poten-cba}
 &&v^2 \lambda_1  = \frac{m_H^2 c_\alpha^2 + m_h^2 s_\alpha^2 - m_{12}^2 t_\beta}{ c_\beta^2}, \ \ \ 
v^2 \lambda_2 = \frac{m_H^2 s_\alpha^2 + m_h^2 c_\alpha^2 - m_{12}^2 t_\beta^{-1}}{s_\beta^2},  \nonumber \\  
&&v^2 \lambda_3 =  \frac{(m_H^2-m_h^2) s_\alpha c_\alpha + 2 m_{H^{\pm}}^2 s_\beta c_\beta - m_{12}^2}{ s_\beta c_\beta }, \ \ \ 
v^2 \lambda_4 = \frac{(\hat{m}_A^2-2m_{H^{\pm}}^2) s_\beta c_\beta + m_{12}^2}{ s_\beta c_\beta },  \nonumber \\
 &&v^2 \lambda_5=  \frac{ - \hat{m}_A^2 s_\beta c_\beta  + m_{12}^2}{ s_\beta c_\beta }\, , 
 \label{eq:lambdas}
\end{eqnarray}
with $\hat{m}_A^2=m_A^2 c_\theta^2+m_X^2 s_\theta^2$.

The general Yukawa interaction is given by
 \bea
&&- {\cal L} =Y_{u2}\,\overline{Q}_L \, \tilde{{ \Phi}}_2 \,u_R
+\,Y_{d2}\,
\overline{Q}_L\,{\Phi}_2 \, d_R\, + \, Y_{\ell 2}\,\overline{L}_L \, {\Phi}_2\,e_R \,\nonumber\\
&&+Y_{u1}\,\overline{Q}_L \, \tilde{{ \Phi}}_1 \,u_R
+\,Y_{d1}\,
\overline{Q}_L\,{\Phi}_1 \, d_R\, + \, Y_{\ell 1}\,\overline{L}_L \, {\Phi}_1\,e_R+\, \mbox{h.c.},
\eea where
$Q_L^T=(u_L\,,d_L)$, $L_L^T=(\nu_L\,,l_L)$,
$\widetilde\Phi_{1,2}=i\tau_2 \Phi_{1,2}^*$, and $Y_{u1,2}$,
$Y_{d1,2}$ and $Y_{\ell 1,2}$ are $3 \times 3$ matrices in family
space. The Yukawa coupling matrices are taken to be aligned to avoid the tree-level flavour changing neutral current \cite{aligned2h,Wang:2013sha},
 \bea
 &&(Y_{u1})_{ii}=\frac{\sqrt{2}m_{ui}}{v}\rho_{1u}, ~~~ (Y_{u2})_{ii}=\frac{\sqrt{2}m_{ui}}{v}\rho_{2u},\nonumber\\
&&(Y_{\ell 1})_{ii}=\frac{\sqrt{2}m_{\ell i}}{v}\rho_{1\ell}, ~~~~ (Y_{\ell 2})_{ii}=\frac{\sqrt{2}m_{\ell i}}{v}\rho_{2\ell},\nonumber\\
&&(X_{d1})_{ii}=\frac{\sqrt{2}m_{di}}{v}\rho_{1d},  ~~~ (X_{d2})_{ii}=\frac{\sqrt{2}m_{di}}{v}\rho_{2d},
\eea
where  $X_{d1,2}=V_{CKM}^\dagger Y_{d1,2} V_{CKM}$, $\rho_{1f}=(c_\beta-s_\beta \kappa_f)$ and $\rho_{2f}=(s_\beta+c_\beta \kappa_f)$ with $f=u,d,\ell$. All the off-diagonal elements are zero.  
The couplings of the neutral Higgs bosons normalized to the SM are given by
\bea\label{hffcoupling} &&
y^{h}_V=\sin(\beta-\alpha),~y_{f}^{h}=\left[\sin(\beta-\alpha)+\cos(\beta-\alpha)\kappa_f\right], \nonumber\\
&&y^{H}_V=\cos(\beta-\alpha),~y_{f}^{H}=\left[\cos(\beta-\alpha)-\sin(\beta-\alpha)\kappa_f\right], \nonumber\\
&&y^{A}_V=0,~y_{A}^{f}=-i\kappa_f~{\rm (for}~u)c_\theta,~y_{f}^{A}=i \kappa_fc_\theta~{\rm (for}~d,~\ell),\nonumber\\ 
&&y^{X}_V=0,~y_{X}^{f}=-i\kappa_f~{\rm (for}~u)s_\theta,~y_{f}^{X}=i \kappa_fs_\theta~{\rm (for}~d,~\ell),
\eea 
where $\alpha$ is the mixing angle of the two CP-even Higgs bosons, and $V$ denotes $Z$ or $W$.

\section{Relevant theoretical and experimental constraints}
In our calculations, we consider the following theoretical and experimental constraints:

{\bf (1) The signal data of the 125 GeV Higgs.} 
We take the light CP even Higgs boson $h$ as the discovered 125 GeV state, and choose $\sin(\beta-\alpha)=1$ to satisfy the bound of the 125 GeV
Higgs signal data, for which the $h$ has the same tree-level couplings to the SM particles as the SM.

{\bf (2) The direct searches and indirect searches for extra Higgses.} 
From the Eq. (\ref{hffcoupling}), one see that the Yukawa couplings of the extra Higgses ($H$, $H^\pm$, $A$, $X$) are proportional 
to $\kappa_u$, $\kappa_d$ and $\kappa_\ell$ for $\sin(\beta-\alpha)=1$. Therefore, we assume $\kappa_u$, $\kappa_d$ and $\kappa_\ell$ to be small enough to  
suppress the production cross sections of these extra Higgses at the LHC, and satisfy the exclusion limits of searches for additional Higgs bosons at the LHC.
Simultaneously, very small $\kappa_u$, $\kappa_d$ and $\kappa_\ell$ can accommodate the indirect
searches for these extra Higgses via the $B$-meson decays. 

{\bf (3) Vaccum stability.} 
We require that the vacuum is stable at tree level, which means that the potential in Eq. (\ref{V2HDM}) has to be bounded from
below and the electroweak vacuum is the global minimum of the full scalar potential. To examine bounded from below condition we 
consider the minimum of quartic part in Eq. (\ref{V2HDM}), $V_{4-min}$, which is written in matrix form in the basis 
$B=\begin{pmatrix} \Phi_1^\dagger \Phi_1 , & \Phi_2^\dagger \Phi_2 , & \chi^2 , & \eta_S^2 \end{pmatrix}^T$,
\begin{align}
    V_{4-min}
    &=C^T \frac{1}{2} \underbrace{\begin{pmatrix}
    \lambda_1 & 
    \lambda_3 + \Delta & 
    \lambda_1' + 2\lambda_4' & 
    \lambda_1' - 2\lambda_4' \\
    \lambda_3 + \Delta & 
    \lambda_2 & 
    \lambda_2' + 2\lambda_5' & 
    \lambda_2' - 2\lambda_5' \\
    \lambda_1' + 2\lambda_4' & 
    \lambda_2' + 2\lambda_5' & 
    \frac{\lambda_1'' + 3\lambda_3''}{6} & 
    \frac{-\lambda_1'' + \lambda_3''}{2} \\
    \lambda_1' - 2\lambda_4' & 
    \lambda_2' - 2\lambda_5' & 
    \frac{-\lambda_1'' + \lambda_3''}{2} & 
    \frac{\lambda_1'' + 3\lambda_3''}{6}
    \end{pmatrix}}_{A} C \nonumber \\
    &= \frac{1}{2} C^T A C ,
    \label{bfbeq}
\end{align}
where $\Delta=0$ for $\lambda_4 \geq |\lambda_5|$ and $\Delta=\lambda_4 - |\lambda_5|$ for $\lambda_4 < |\lambda_5|$.

A copositive matrix $A$ is required to ensure the potential to be bounded from below.
Following the approaches described in \cite{Kannike:2012pe,Dutta:2023cig}, the matrix $A$ need satisfy
 $\det(A) \geq 0  \lor  (\text{adj} A)_{ij} < 0$, for some $i,j$. 
 The adjugate of $A$ is the transpose of the cofactor matrix of $A$: $(\text{adj} A)_{ij} = (-1)^{i+j}M_{ji}$, with $M_{ij}$ being the determinant of 
the submatrix that results from deleting row $i$ and column $j$ of $A$.
In addition, one deletes the $i$-th row and
column of $A$, $i=1,2,3,4$, and obtains 4 symmetric $3\times 3$ matrices, which are
 required to be copositive. The copositivity of the symmetric order 3 matrix $B$ with entries $b_{ij}$, $i,j=1,2,3$ requires
    \begin{align}
        &b_{11} \geq 0 , \quad b_{22} \geq 0 , \quad b_{33} \geq 0,  \nonumber\\ 
        &\Bar{b}_{12} = b_{12} + \sqrt{b_{11}b_{22}} \geq 0,  \nonumber\\
        &\Bar{b}_{13} = b_{13} + \sqrt{b_{11}b_{33}} \geq 0,  \nonumber\\
        &\Bar{b}_{23} = b_{23} + \sqrt{b_{22}b_{33}} \geq 0, \nonumber\\
        &\sqrt{b_{11}b_{22}b_{33}} + b_{12}\sqrt{b_{33}} + b_{13}\sqrt{b_{22}} + b_{23}\sqrt{b_{11}} + \sqrt{2 \Bar{b}_{12} \Bar{b}_{13} \Bar{b}_{23}} \geq 0. \label{bfb5}
    \end{align}

{\bf (4) Tree-level perturbative unitarity.} 
We demand that the amplitudes of the scalar quartic interactions leading to $2\to 2$ scattering
processes remain below the value of $8\pi$ at tree-level, which is implemented in $\textsf{SPheno-v4.0.5}$ \cite{Porod:2003um} using $\textsf{SARAH-SPheno files}$ \cite{Staub:2013tta}.

{\bf (5) The oblique parameters.}
The oblique parameters ($S$, $T$, $U$) can obtain additional corrections via 
the self-energy diagrams exchanging $H$, $H^\pm$, $A$, and $X$. For $\sin(\beta-\alpha)=1$, the expressions of $S$, $T$ and $U$ in the 
 model are approximately written as \cite{stu1,stu2}
\bea\label{stu-eq}
S&=&\frac{1}{\pi m_Z^2}
\left[  c_\theta^2 F_S(m_Z^2,m_H^2,m_A^2) + s_\theta^2 F_S(m_Z^2,m_H^2,m_X^2)
-F_S(m_Z^2,m_{H^{\pm}}^2,m_{H^{\pm}}^2) \right], \nonumber \\
T&=&\frac{1}{16\pi m_W^2 s_W^2} \left[ - c_\theta^2 F_T(m_H^2,m_A^2) - s_\theta^2 F_T(m_H^2,m_X^2)
+ F_T(m_{H^{\pm}}^2,m_H^2) \right. \nonumber \\
&&+ \left. c_\theta^2 F_T(m_{H^{\pm}}^2,m_A^2) + s_\theta^2 F_T(m_{H^{\pm}}^2,m_X^2)\right], \nonumber\\
U&=&\frac{1}{\pi m_W^2} \left[ F_S(m_W^2,m_{H^{\pm}}^2,m_H^2) -2 F_S(m_W^2,m_{H^{\pm}}^2,m_{H^{\pm}}^2) \right.\nonumber\\
&& \left.+c_\theta^2 F_S(m_W^2,m_{H^{\pm}}^2,m_A^2) + s_\theta^2 F_S(m_W^2,m_{H^{\pm}}^2,m_X^2) \right]\nonumber\\
&&-\frac{1}{\pi m_Z^2} \left[  c_\theta^2 F_S(m_Z^2,m_H^2,m_A^2) + s_\theta^2 F_S(m_Z^2,m_H^2,m_X^2) - F_S(m_Z^2,m_{H^{\pm}}^2,m_{H^{\pm}}^2) \right],
\eea
where
\beq\label{stu-ft}
F_T(a,b)=\frac{1}{2}(a+b)-\frac{ab}{a-b}\log(\frac{a}{b}),~~F_S(a,b,c)=B_{22}(a,b,c)-B_{22}(0,b,c),
\eeq
with
\bea
&&B_{22}(a,b,c)=\frac{1}{4}\left[b+c-\frac{1}{3}a\right] - \frac{1}{2}\int^1_0 dx~X\log(X-i\epsilon),\nonumber\\
&&
X=bx+c(1-x)-ax(1-x).
\eea
Ref. \cite{pdg2020} gave the fit results of $S$, $T$ and $U$,
\beq
S=-0.01\pm 0.10,~~  T=0.03\pm 0.12,~~ U=0.02 \pm 0.11, 
\eeq
with the correlation coefficients 
\beq
\rho_{ST} = 0.92,~~  \rho_{SU} = -0.80,~~  \rho_{TU} = -0.93.
\eeq

\begin{figure}[tb]
\centering
\includegraphics[width=5.cm]{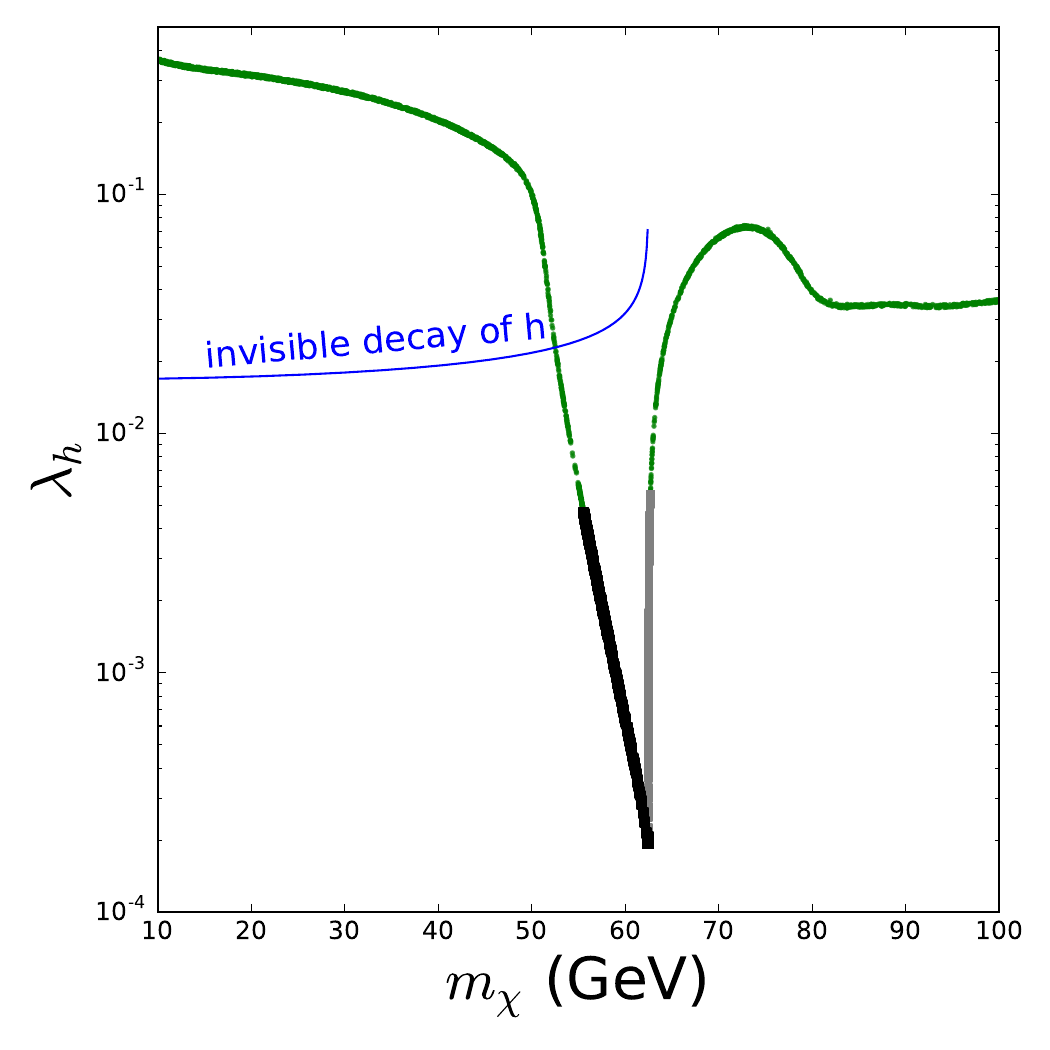}
\vspace{-0.4cm}\caption{$\lambda_h$ consistent with the relic data versus $m_\chi$. The dark thick line is allowed by the direct and indirect searches for DM, and the light thick line is excluded by the indirect searches.}
\label{dmrelic}
\end{figure}

\section{Dark matter} 
The two neutral CP-even Higgs can mediate the interactions of DM, $\lambda_h v \chi^2 h/2$ and $\lambda_H v \chi^2 H/2$ with
\bea
\lambda_h  &\equiv& (\lambda_2^\prime + 2\lambda_5^\prime) v s_\beta c_\alpha - (\lambda_1^\prime + 2\lambda_4^\prime) v c_\beta s_\alpha,\nonumber\\
\lambda_H  &\equiv& (\lambda_2^\prime + 2\lambda_5^\prime) v s_\beta s_\alpha + (\lambda_1^\prime + 2\lambda_4^\prime) v c_\beta c_\alpha.
\eea
We consider a light DM whose freeze-out temperature is much lower than that of EWPT, and thus the effect of EWPT on the current DM relic density can be ignored.
We take the new scalars to be much heavier than the DM so that the DM pair-annihilation to these new scalars are kinematically forbidden.
In the parameter space chosen previously, the couplings of $H$ to the SM particles can be ignored. Therefore, the DM relic density 
hardly constrains the $\lambda_H$, and $\lambda_{1,2,4,5}^\prime$ are allowed to have room enough to produce the pattern of EWPT
needed by the EWBG. The annihilation processes with $s$-channel exchange of $h$ are responsible for the relic density.
However, for a light $\chi$, the invisible decay $h\to \chi\chi$ is kinematically allowed, and the signal data of the 125 GeV Higgs impose 
strong upper limits on the $h\chi\chi$ coupling \cite{invisible}, which is possible to conflict with the requirement of the correct relic density
The elastic scattering of $\chi$ on a nucleon receives the contributions of the process with $t$-channel exchange of $h$, which
can be strongly constrained by the direct searches experiments of XENON \cite{XENON:2018voc}. Besides, the indirect searches for DM
can impose upper limits on the averaged cross sections
of the DM annihilation to $e^+ e^-$, $\mu^+ \mu^-$, $\tau^+\tau^-$, $u\bar{u}$, $b\bar{b}$, and $WW$ \cite{Fermi-LAT:2015att}. 

After imposing the relevant theoretical and experimental constraints mentioned previously, we show
the $\lambda_h$ versus $m_\chi$ allowed by the 
invisible decay of the 125 GeV Higgs, the DM relic density, the direct and indirect searches experiments in Fig. \ref{dmrelic}.
From Fig. \ref{dmrelic}, we find that the DM with a mass of $55$ GeV $\sim 62.5$ GeV is allowed by the joint constraints of
the invisible decay of the 125 GeV Higgs, the DM relic density, the direct and indirect searches experiments.

\section{EWPT and Baryogenesis}

We first consider the effective scalar potential at the finite temperature. The
neutral elements of $\Phi_1$ and $\Phi_2$ are shifted by $\frac{h_1}{\sqrt{2}}$ and $\frac{h_2+ih_3}{\sqrt{2}}$.
It is plausible to take the imaginary part of the neutral elements of $\Phi_1$ to be zero since the effective potential of Eq. (\ref{V2HDM}) only depends on the relative phase of the
neutral elements of $\Phi_1$ and $\Phi_2$ .

The complete effective potential at finite temperature contains the tree level potential, the
Coleman-Weinberg term \cite{vcw}, the finite temperature corrections \cite{vloop} and the resummed daisy corrections \cite{vring1,vring2}, which is gauge-dependent \cite{vgauge1,vgauge2}. 
Here we consider the high-temperature approximation of effective potential, which keeps only the thermal mass terms in the high-temperature 
expansion and the tree-level potential. Therefore, the effective potential is gauge invariant, and it does not depend on the renormalization scheme and the
resummation scheme.
The high-temperature approximation of effective potential is given by
\begin{align}\label{veff0}
&V_{eff}(h_1,h_2,h_3,\chi,\eta_s) =  \frac{1}{2} (m_{11}^2 + \Pi_{h_1}) h_1^2+ \frac{1}{2} (m_{22}^2+ \Pi_{h_2}) (h_2^2 + h_3^2)+ \frac{1}{2} (m_{S}^2 + {m^\prime_S}^{2} +\Pi_{\chi}) \chi^2
\nonumber\\
&+ \frac{1}{2} (m_{S}^2-{{m^\prime_S}^{2}} +\Pi_{\eta_s}) \eta_s^2 
+ \frac{1}{8}(\lambda_1h_1^4+\lambda_2h_2^4 +\lambda_2 h_3^4)+\frac{1}{4} \lambda_{345}h_1^2h_2^2 +\frac{1}{4} \bar{\lambda}_{345} h_1^2 h_3^2\nonumber\\
&+\frac{\lambda_2}{4}h_3^2h_2^2-m_{12}^2h_1h_2-\frac{\mu}{\sqrt{2}}h_3\eta_sh_1
+\frac{\lambda^{\prime}_1}{4}(\chi^2+\eta_s^2)h_1^2+\frac{\lambda^{\prime}_2}{4}(\chi^2+\eta_s^2)(h_2^2+h_3^2) \nonumber\\
&+ \frac{\lambda^{\prime}_4}{2}(\chi^2-\eta_s^2)h_1^2 +\frac{\lambda^{\prime}_5}{2}(\chi^2-\eta_s^2)( h_3^2+h_2^2)
+ (\frac{\lambda^{''}_1}{48}+\frac{\lambda^{\prime\prime}_3}{16}) (\chi^4+\eta_s^4)+\frac{1}{8}(\lambda^{\prime\prime}_3-\lambda^{\prime\prime}_1)\chi^2\eta_s^2,\nonumber\\
&\Pi_{h_1}= \left[{9g^2\over 2} + {3g'^2\over 2} + 6\lam_{1} +4\lam_{3} +2\lam_4 + 2\lambda^\prime_{1} + 6y_t^2 c_\beta^2\right] {T^2 \over 24},\nonumber\\
&\Pi_{h_2}= \left[{9g^2\over 2} + {3g'^2\over 2} + 6\lam_{2} +4\lam_{3} +2\lam_4 + 2\lambda^\prime_{2} + {6y_t^2 s_\beta^2}\right] {T^2 \over 24},\nonumber\\
&\Pi_{h_3} =\Pi_{h_2},\nonumber\\
&\Pi_{\chi}= \left[ 4\lambda^\prime_{1} + 4\lambda^\prime_{2} +8\lambda^\prime_{4} + 8\lambda^\prime_{5} + 2\lambda^{\prime\prime}_3 \right] {T^2 \over 24},\nonumber\\
&\Pi_{\eta_s}=  \left[ 4\lambda^\prime_{1} + 4\lambda^\prime_{2} - 8\lambda^\prime_{4} - 8\lambda^\prime_{5} + 2\lambda^{\prime\prime}_3 \right] {T^2 \over 24},
\end{align}
where $\bar{\lambda}_{345}=\lambda_3+\lambda_4-\lambda_5$, $y_t={\sqrt{2} m_t \over v}$, and $\Pi_{i}$ denotes the thermal mass terms of the field $i$.

Because baryogenesis is driven by diffusion processes in front of the bubble wall, one needs to
compute the $T_n$ at which bubble nucleation actually starts. This can be calculated 
straightforwardly from the nucleation rate per unit volume by \cite{bubble-0,bubble-1,bubble-2},
$\Gamma \approx A(T)e^{-S_3/T}$,
where $A(T)\sim T^4$ is a prefactor and $S_3$ is a three-dimensional Euclidian action.
The nucleation temperature $T_n$ is obtained by $\Gamma/H^4$= 1, where $H$ is the Hubble parameter.
It is roughly estimated by $\frac{S_3(T)}{T}|_{T = T_n}= 140$. 
The bubble wall VEV profiles can be determined by
the bounce equations of fields.

The WKB approximation method is used to evaluate the CP-violating source terms and chemical potentials transport equations of particle species in the wall frame with 
a radial coordinate $z$ \cite{bg2h-3,0006119,0604159}.
A top quark penetrating the bubble wall acquires a complex mass as a function of $z$,
\bea
&&m_t(z)=\frac{y_t}{\sqrt{2}}\sqrt{(c_\beta h_1(z)+s_\beta h_2(z))^2+s_\beta^2 h^2_3(z)}~ e^{i\Theta_t(z)},\nonumber\\
&&{\rm with~}\Theta_t(z)=\varphi_Z(z)+\arctan\frac{s_\beta h_3(z)}{c_\beta h_1(z)+s_\beta h_2(z)},\nonumber\\
&&\partial_z\varphi_Z(z)=-\frac{h^2_2(z)+h^2_3(z)}{h^2_1(z)+h^2_2(z)+h^2_3(z)}\partial_z\varphi_(z),\nonumber\\
&&\varphi_(z)=\arctan\frac{h_3(z)}{h_2(z)}.
\eea 
 In our calculation, the imaginary part of the neutral element of $\Phi_1$ is taken to be zero. As a result, 
 the nonvanishing $Z_\mu$ field induces an additional CP-violating force acting on the top quark, which is removed a local axial transformation of top quark,
reintroducing an additional overall phase $\varphi_Z(z)$ into $m_t$ \cite{bg2h-4}.

The transport equations are derived by the complex mass of the top quark,
and contains effects of the strong sphaleron process ($\Gamma_{ss}$) \cite{bg2h-3,9311367}, W-scattering ($\Gamma_W$) \cite{bg2h-3,9506477}, the top Yukawa interaction ($\Gamma_y$) \cite{bg2h-3,9506477},
 the top helicity flips ($\Gamma_M$) \cite{bg2h-3,9506477}, and the Higgs number violation ($\Gamma_h$) \cite{bg2h-3,9506477}. The transport equations are written as
	\begin{align}
		0 =   & 3 v_W K_{1,t} \left( \partial_z \mu_{t,2} \right) + 3v_W K_{2,t} \left( \partial_z m_t^2 \right) \mu_{t,2} + 3 \left( \partial_z u_{t,2} \right) \notag
		\\ &- 3\Gamma_y \left(\mu_{t,2} + \mu_{t^c,2} + \mu_{h,2} \right) - 6\Gamma_M \left( \mu_{t,2} + \mu_{t^c,2} \right) - 3\Gamma_W \left( \mu_{t,2} - \mu_{b,2} \right) \notag
		\\ &- 3\Gamma_{ss} \left[ \left(1+9 K_{1,t} \right) \mu_{t,2} + \left(1+9 K_{1,b} \right) \mu_{b,2} + \left(1-9 K_{1,t} \right) \mu_{t^c,2} \right] \notag \,,\\
                0=    & 3 v_W K_{1,t} \left( \partial_z \mu_{t^c,2} \right)  + 3v_W K_{2,t} \left( \partial_z m_t^2 \right)  \mu_{t^c,2} + 3 \left( \partial_z u_{t^c,2} \right) \notag   \\
		      & - 3\Gamma_y \left(\mu_{t,2} + \mu_{b,2} + 2\mu_{t^c,2} + 2\mu_{h,2} \right) - 6\Gamma_M \left( \mu_{t,2} + \mu_{t^c,2} \right) \notag    \\
		      & - 3\Gamma_{ss} \left[ \left( 1+9 K_{1,t}\right) \mu_{t,2} + \left(1+9K_{1,b}\right) \mu_{b,2} + \left(1-9K_{1,t}\right) \mu_{t^c,2} \right]  \notag \,,          \\
		0 =   & 3v_W K_{1,b} \left(\partial_z \mu_{b,2}\right) + 3 \left(\partial_z u_{b,2} \right) - 3\Gamma_y \left( \mu_{b,2} + \mu_{t^c,2} + \mu_{h,2} \right) - 3\Gamma_W \left( \mu_{b,2} - \mu_{t,2} \right) \notag \,,   \\
		      & - 3\Gamma_{ss} \left[ \left( 1 + 9K_{1,t}\right) \mu_{t,2} + (1+9K_{1,b}) \mu_{b,2} + (1-9K_{1,t}) \mu_{t^c,2} \right] \notag \,, \\
		0 =   & 4v_W K_{1,h} \left( \partial_z \mu_{h,2}\right) +
		4\left( \partial_z u_{h,2}\right) - 3\Gamma_y \left(
		\mu_{t,2} + \mu_{b,2} + 2\mu_{t^c,2} + 2\mu_{h,2} \right) -
		4\Gamma_h
		\mu_{h,2}  \notag\,,\nonumber
\end{align}
\begin{align}
		S_t = & -3K_{4,t} \left( \partial_z \mu_{t,2}\right) + 3v_W \tilde{K}_{5,t} \left( \partial_z u_{t,2}\right) + 3v_W \tilde{K}_{6,t} \left( \partial_z m_t^2 \right) u_{t,2} + 3\Gamma_t^\mathrm{tot} u_{t,2} \notag \,,       \\
		0 =   & -3K_{4,b} \left( \partial_z \mu_{b,2} \right) + 3v_W \tilde{K}_{5,b} \left(\partial_z u_{b,2}\right) + 3\Gamma_b^\mathrm{tot} u_{b,2}  \notag  \,,   \\
		S_t = & -3K_{4,t} \left( \partial_z \mu_{t^c,2}\right) + 3v_W \tilde{K}_{5,t} \left( \partial_z u_{t^c,2}\right) + 3v_W \tilde{K}_{6,t} \left( \partial_z m_t^2\right) u_{t^c,2} + 3\Gamma_t^\mathrm{tot} u_{t^c,2} \notag \,, \\
		0 =   & -4K_{4,h} \left( \partial_z \mu_{h,2} \right) + 4v_W \tilde{K}_{5,h} \left( \partial_z u_{h,2} \right) + 4\Gamma_h^\mathrm{tot} u_{h,2}  \,,
\label{TransportEquations}
\end{align}

The $\mu_{i,2}$ and $u_{i,2}$ are the second-order CP-odd chemical potential and the
plasma velocity of the particle $i=t,~t^c,~b,~h$,. The source term $S_t$ is 
\beq \label{Eq:TransportEquations:Source}
	S_t = -v_W K_{8,t} \partial_z \left( m_t^2 \partial_z \theta_t \right) + v_W K_{9,t} \left( \partial_z \theta_t \right) m_t^2  \left( \partial_z m_t^2\right).
\eeq
The functions $K_{a,i}$ and $\tilde{K}_{a,i}$ ($a=1-9$) are defined in Ref. \cite{0604159}, and
the $\Gamma_{i}^{\mathrm{tot}}$ are the total reaction rate of the particle $i$ \cite{bg2h-3,0604159}.

The chemical potential of the left-handed quarks $\mu_{B_L}$ is obtained by solving the transport equations.
The left-handed quark number is converted into a baryon asymmetry by the weak sphalerons, which is calculated as 
\begin{equation}
	Y_B = \frac{405 \Gamma_{ws}}{4\pi^2 v_w g_* T_{n}}\int_0^\infty dz \mu_{B_L}(z) f_{sph}(z) \exp\left(-\frac{45\Gamma_{ws}z}{4v_w}\right)\,,
\end{equation}
where $\Gamma_{ws}\simeq 10^{-6} T_{n}$
is the weak sphaleron rate inside bubble \cite{sphaleron-ws} and the wall velocity $v_w$ is taken as 0.1. 
The function $f_{sph}(z) = min(1, 2.4 \frac{T_n}{\Gamma_{ws}} e^{-40\xi_n(z)/T_n} )$ with $\xi_n(z)=\sqrt{\langle h_1\rangle^2+\langle h_2\rangle^2+\langle h_3\rangle^2}$ 
is used to smoothly interpolate between the sphaleron
rates in the broken and unbroken phases \cite{bg2h-4}.

\begin{table}[t]
\centering
\begin{tabular}{ | c | c |c | c | c | c |c | c | c | c |c | c | c | c | c | c | c |c | c | c | c |c |}
\hline
 $m_h$(GeV) &$m_{H}=m_{H^\pm}$(GeV)& $m_{\chi}$(GeV) & $m_{A}$(GeV) & $m_{X}$(GeV)& $m_{12}^2$(GeV$)^2$  \\
\hline
  125.0  & 467.69 &  55.95  &  69.80  & 333.67 & 2740.09 \\
 \hline
\end{tabular}
\centering
\begin{tabular}{ | c | c| c | c | c |c | c | c | c |c | c | c | c |c | c | c | c | c | c | c |c | c | c | c |}
\hline
$t_\beta$& $\sin(\beta-\alpha)$ &$\sin\theta$ & $\lambda^{\prime}_1$  & $\lambda^{\prime}_2$  & $\lambda^{\prime}_4$  & $\lambda^{\prime}_5$  & $\lambda^{\prime\prime}_1=\lambda^{\prime\prime}_3$ \\
\hline
~1.0~ &~1.0~ & 0.324 & 2.293 & 1.351& -1.143 & -0.675 & 1.839 \\
 \hline
\end{tabular}
\caption{Input parameters for the BP1, and other parameters are given above.}
\label{tabbp1}
\end{table}

\begin{figure}[tb]
\centering
\includegraphics[width=11.cm]{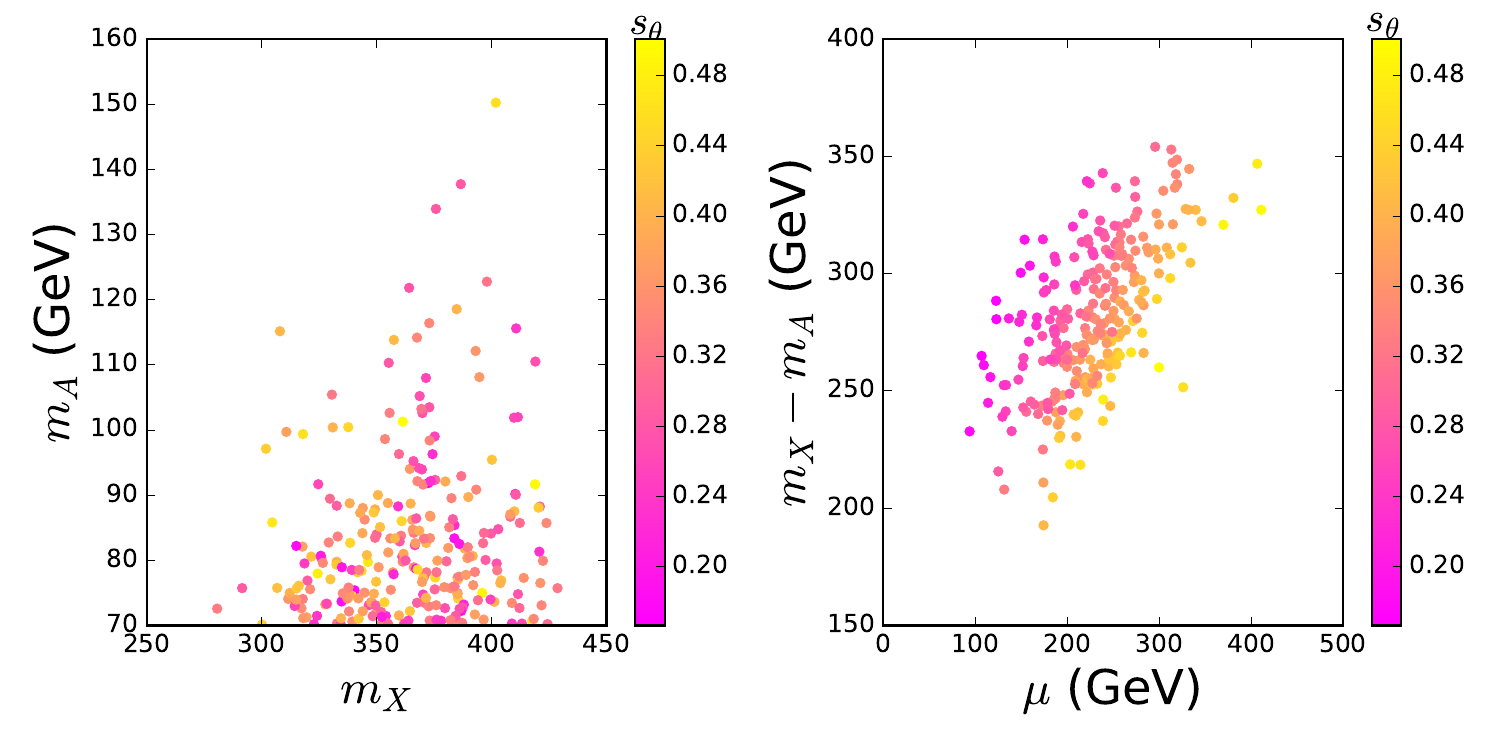}
\vspace{-0.4cm}\caption{The scattering plots achieving the three-step PTs with the characteristics mentioned in the text, where we take $m_H=m_{H^\pm}$ and $0.1<s_\theta<0.7$.}
\label{figscan}
\end{figure}

We focus on the following three-step PTs achieving the spontaneous CP-violation at high temperature and recovering the CP-symmetry. At the first-step PT, the $\eta_S$ field firstly acquires a nonzero
VEV while $h_{1,2,3}$ still remains zero. The second-step PT is a strongly first-order EWPT converting the origin phase of $h_{1,2,3}$
into an electroweak symmetry broken phase, where $h_{3}$ is required to be nonzero. In order to prevent the electroweak sphalerons to
wash out the produced BAU inside the bubbles of broken
phase, the PT strength is impose an bound \cite{pt-stren},
$\frac{\xi_n}{T_n} > 1.0$ in the broken phase.
After the third-step PT, the observed vacuum is produced and the CP-symmetry is restored while the BAU is not changed.
We employ the package $\textsf{CosmoTransitions}$ to analyze the PTs \cite{cosmopt}.
Some parameter space achieving the three-step PTs are shown in Fig. \ref{figscan}, where we consider the constraints of the vacuum stability, oblique parameter \cite{pdg2020}, 
dark matter observables, and the 125 GeV Higgs signal data, and the data of BAU is not included.
From Fig. \ref{figscan}, we find that the three-step PTs satisfying our requirements favor an appropriate value of $\mu$ since the $\mu$ term of Eq. (\ref{veff0}) can lead
to a close correlation between $\langle h_3\rangle$ and $\langle \eta_s \rangle$ of the potential minimum. 
As a result, according to Eq. (\ref{eq:mum0}), $m_A$ and $m_X$ is required to have a large mass splitting.



\begin{figure}[tb]
\centering
\includegraphics[width=11.cm]{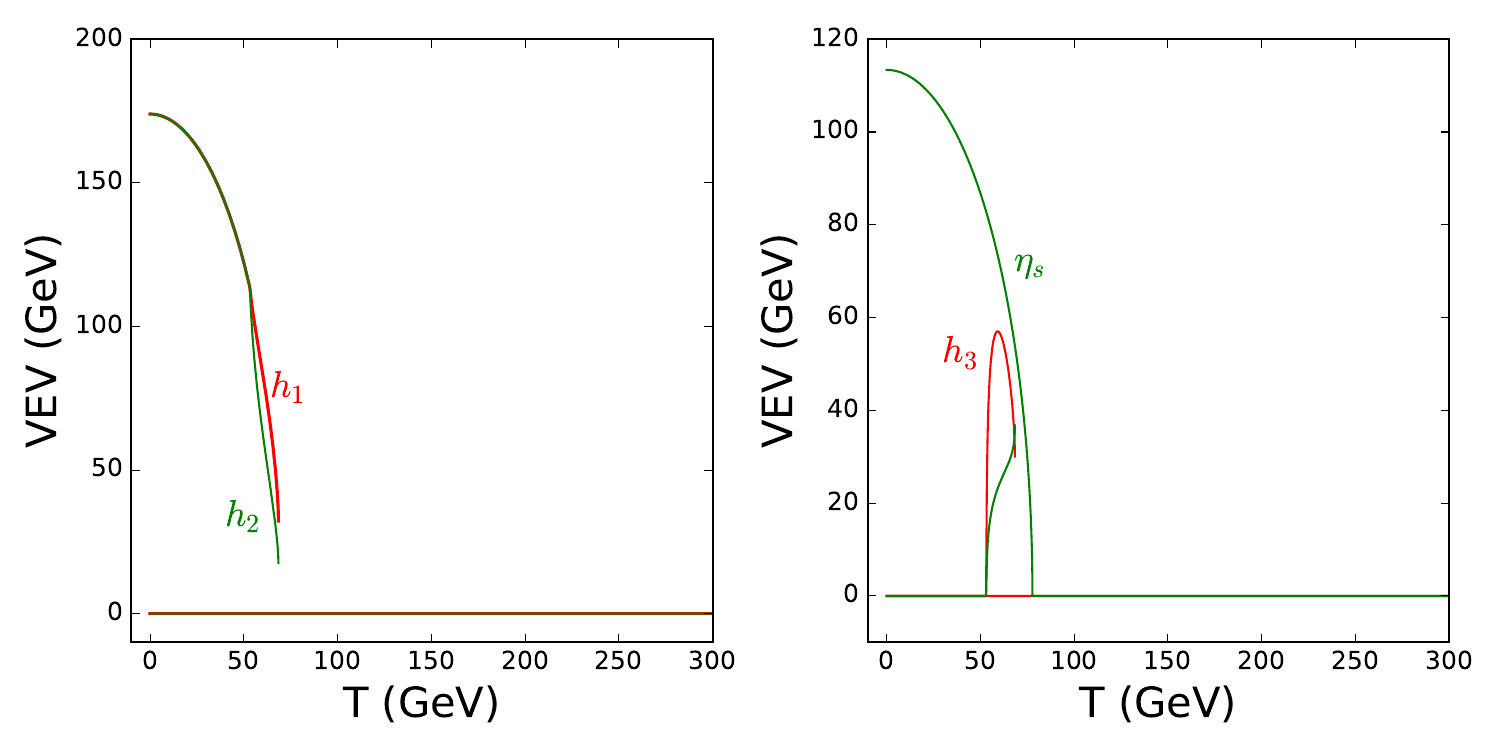}
\vspace{-0.4cm}\caption{The phase histories of the BP1, where $\langle \chi \rangle$ is always 0.}
\label{figpt}
\end{figure}

\begin{figure}[tb]
\centering
\includegraphics[width=7.cm]{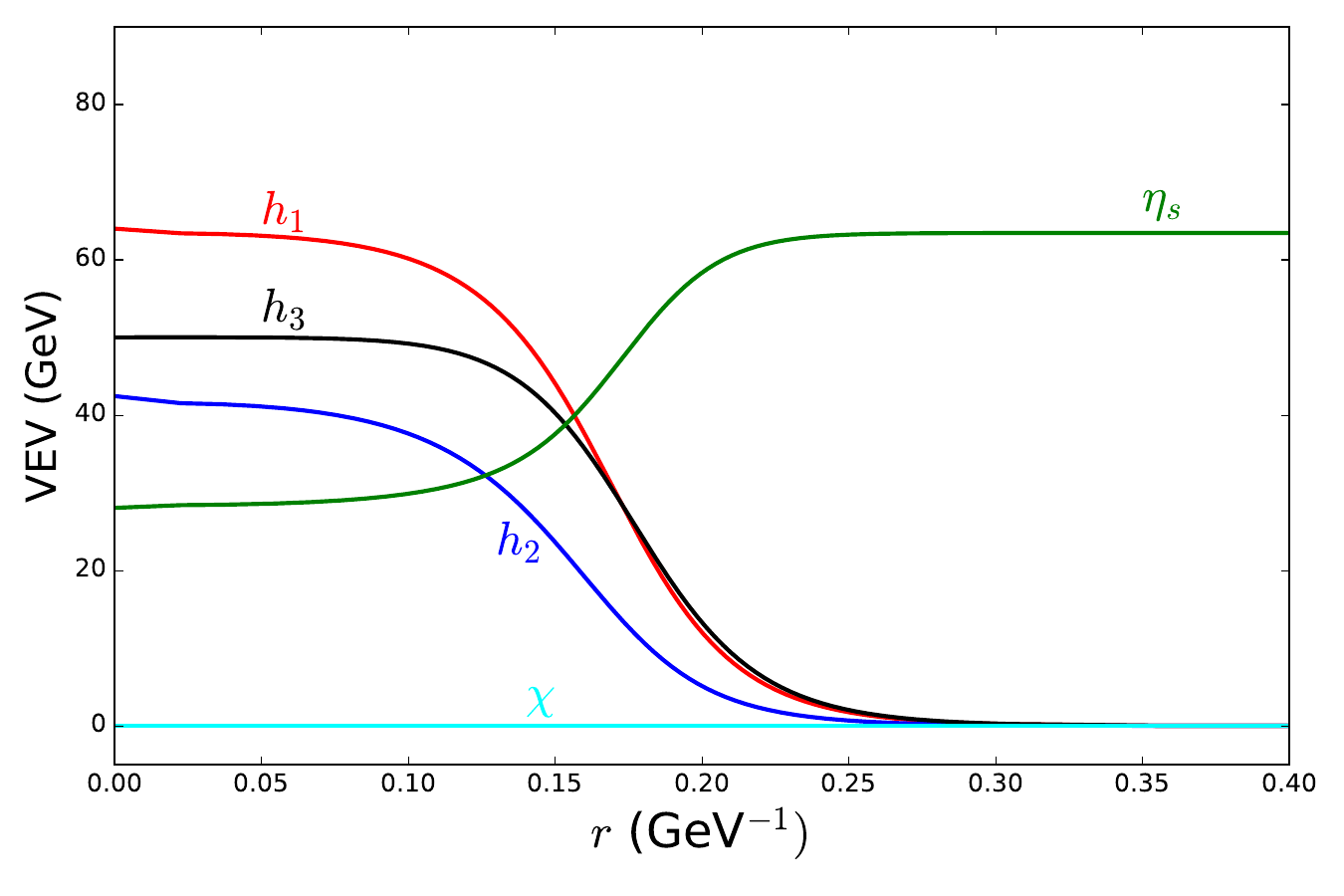}
\vspace{-0.4cm}\caption{The radial nucleation bubble wall VEV profiles for the first-order EWPT.}
\label{figbounce}
\end{figure}

\begin{figure}[tb]
\centering
\includegraphics[width=7.cm]{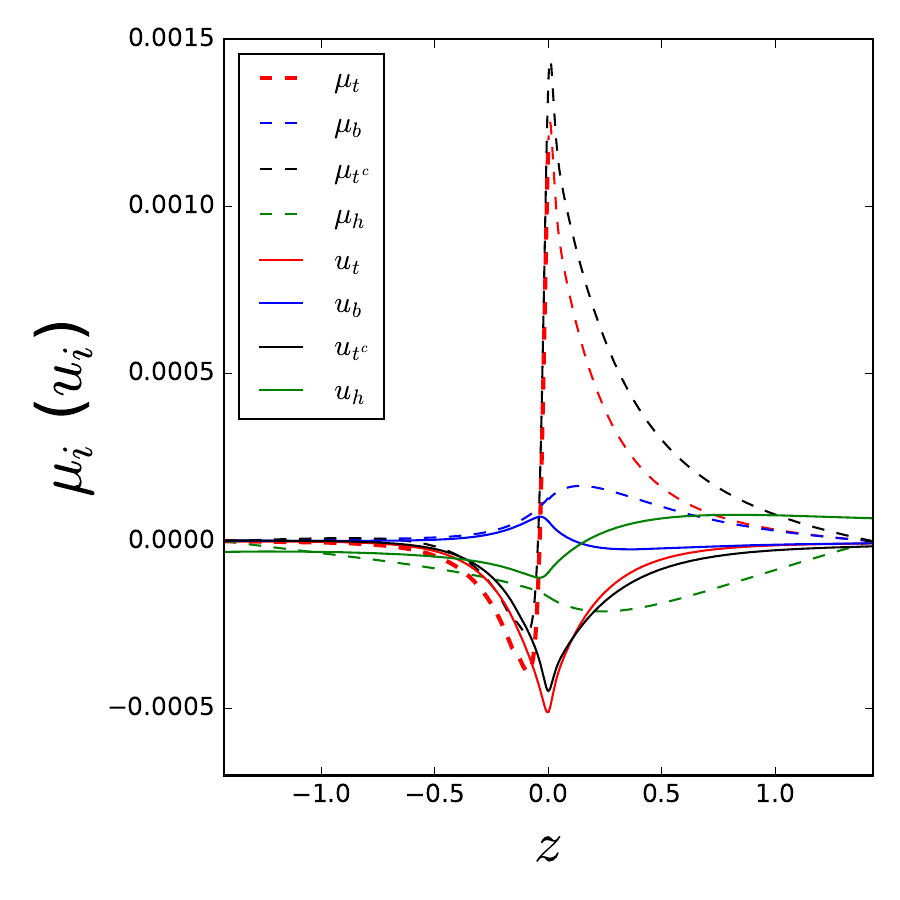}
\vspace{-0.4cm}\caption{The $\mu_i$ and $u_i$ from the transport equations as
functions of the position of the bubble wall.}
\label{figmu}
\end{figure}

We pick out a benchmark point BP1 to discuss the EWPT and baryogenesis detailedly, and the key input parameters are
shown in Table \ref{tabbp1}. The phase histories for the BP1 are shown in Fig. \ref{figpt}.
Because the contributions of the thermal mass terms to the effective potential are proportional to $T^2$,
the minimum of the potential is at the origin at a very high temperature.
As the Universe cools, at T=85.38 GeV, a second-order PT takes place during which 
$\eta_s$ acquires a nonzero VEV and the other four fields remain zero.
At T=69.65 GeV, a strongly first-order EWPT starts which breaks electroweak symmetry,  (0, 0, 0, 0, 73.71) GeV $\to$ (62.42, 34.64, 55.24, 0.0, 37.50) GeV
 for ($\langle h_1\rangle$, $\langle h_2\rangle$, $\langle h_3\rangle$, $\langle\chi\rangle$, $\langle\eta_s\rangle$). 
The PT strength is 1.30, and the BAU is produced via the EWBG mechanism.
At T=52.95 GeV, another second-order PT happens, and then CP-symmetry is recovered. The vacuum evolves along the final phase, and
 ultimately ends in the observed values at T = 0 GeV. Meanwhile, $\xi_{n}>1$ is always kept so that the BAU is not washed out by
the sphaleron processes.
The freeze-out temperature of $\chi$ with a mass of 55.95 GeV is around 2.8 GeV, which is much lower than the PT temperatures.

The calculation of BAU depends on the bubble wall profiles, and we use the $\textsf{FindBounce}$ \cite{findbounce} to obtain the bubble wall VEV profiles for the first-order EWPT of BP1, which is given 
in Fig. \ref{figbounce}.  The WKB method of calculating transport equations needs the condition of $L_W T_n \gg 1$, where $L_W$ is the width of bubble wall.
The $L_W T_n$ of BP1 is approximately estimated to be 3.4.

\section{Comment on gravitational wave signatures, dark matter, and the LHC signatures}
 
The first-order EWPT needed by the EWBG can produce the gravitational wave. 
We find that the gravitational wave signatures from the three-step PTs mentioned above can easily exceed the sensitivity curve of the U-DECIGO detector \cite{udecigo}, such as those of BP1.
A full exploration of the parameter space will potentially find promising regions for detectable gravitational wave signal at the BBO \cite{bbodecigo}.
The extra Higgses ($H$, $H^\pm$, $A$, $X$) couplings to the SM fermions are significantly suppressed for $\kappa_{u,d,\ell}\to 0$.
Therefore, these extra Higgses are dominantly produced at the LHC via the electroweak processes mediated by $Z,W^\pm$, and $\gamma$,
and the main decay modes include $H\to AZ$, $H^\pm \to A W^\pm$. The $A$ decay modes depend on specific values of $\kappa_{u,d,\ell}$.
For a heavy DM whose freeze-out temperature is higher than the EWPT temperature, the EWPT can
give significant effects on the DM relic density. 
The studies of the LHC signatures and the heavy DM will be carried out in the future.

\section{Conclusion}

 We proposed a complex singlet scalar extension of the 2HDM respecting a discrete dark CP-symmetry.
The dark CP-symmetry guarantees $\chi$ to be a DM candidate on one hand and on the other hand allows $\eta_s$ to have mixings with 
the pseudoscalars of the scalar doublet fields, which plays key roles in producing the CP-violation sources needed by
the EWBG at high temperature.
Imposing relevant theoretical and experimental constraints, we studied the scenario of
$m_{\chi}$ around the SM Higgs resonance region, and found that the dark matter relic abundance and the BAU can be simultaneously explained.

\section*{Acknowledgment}
This work was supported by the Natural Science Foundation of
Shandong province ZR2023MA038, andthe National Natural Science Foundation
of China under grant 11975013. 



\end{document}